\documentclass[prl,superscriptaddress,twocolumn]{revtex4}
\usepackage{amsmath}
\usepackage{amsfonts}
\usepackage{graphicx}
\usepackage{hyperref}
\usepackage{braket}
\usepackage{gensymb}

\begin{document}

\title{Stable Weyl points, trivial surface states and particle-hole compensation in WP$_2$}

\author{E. Razzoli}
\email{razzoli@physics.ubc.ca}
\author{B. Zwartsenberg}
\author{M. Michiardi}
\author{F. Boschini}
\author{R. P. Day}
\author{I. S. Elfimov}
\affiliation{Quantum Matter Institute, University of British Columbia, Vancouver, BC V6T 1Z4, Canada}
\affiliation{Department of Physics and Astronomy, University of British Columbia, Vancouver, BC V6T 1Z1, Canada}
\author{J. D. Denlinger}
\affiliation{Advanced Light Source, Lawrence Berkeley National Laboratory, California 94720, USA}
\author{V. S\"uss}
\author{C. Felser}
\affiliation{Max Planck Institute for Chemical Physics of Solids, 01187 Dresden, Germany}
\author{A. Damascelli}
\email{damascelli@physics.ubc.ca}
\affiliation{Quantum Matter Institute, University of British Columbia, Vancouver, BC V6T 1Z4, Canada}
\affiliation{Department of Physics and Astronomy, University of British Columbia, Vancouver, BC V6T 1Z1, Canada}

\date{\today}

\begin{abstract}

A possible connection between extremely large magneto-resistance and the presence of Weyl points has garnered much attention in the study of topological semimetals. Exploration of these concepts in transition metal phosphide WP$_2$ has been complicated by conflicting experimental reports. Here we combine angle-resolved photoemission spectroscopy (ARPES) and density functional theory (DFT) calculations to disentangle surface and bulk contributions to the ARPES intensity, the superposition of which has plagued the determination of the electronic structure in WP$_2$.  Our results show that while the hole- and electron-like Fermi surface sheets originating from surface states have different areas, the bulk-band structure of WP$_2$ is electron-hole-compensated in agreement with DFT. Furthermore, the detailed band structure is compatible with the presence of at least 4 temperature-independent Weyl points, confirming the topological nature of WP$_2$ and its stability against lattice distortions.

\end{abstract}

\maketitle

Since the first observation of topological Fermi arcs in TaAs \cite{Xu2015},  Weyl semimetals have been the subject of much interest. Recently, it was suggested that non-centrosymmetric transition metal dichalcogenides (TMDs) such as WTe$_2$, host type-II of Weyl points (WPs) \cite{Soluyanov2015}. This classification refers to those Weyl-semimetals wherein tilted three-dimensional cones form at the intersection of an electron and hole pocket.
Beyond the presence of such WPs, semimetallic TMDs also display an  extremely large magnetoresistance (XMR), which scales with the square of the applied magnetic field \cite{Ali2014, Chen2016}. 
Motivated by the semiclassical theory of transport \cite{AshcroftMermin}, particle-hole compensation was suggested as the source of this XMR signal \cite{Ali2014, Pletikosi2014}. In this scenario, changes in the band structure that remove the particle-hole compensation, as observed for instance in  WTe$_2$, would eliminate the possibility of XMR. The viability of this interpretation is further reinforced by its ability to explain the strong temperature dependence of the XMR, i.e. its so-called `turn-on behaviour' \cite{Pletikosi2014, Wu2015, Thoutam2015}. 
However, the observation of XMR in an uncompensated system such as MoTe$_2$ challenges this interpretation \cite{Thirupathaiah2017}.
In light of this, magnetic field-induced modifications to the electronic band structure, such as the formation of excitonic \cite{Khveshchenko2001} or charge density wave gaps \cite{Trescher2017}, have been proposed as alternative explanations. To date, no consensus has been established regarding the origin of the XMR in type-II Weyl semimetals.

More recently, density functional theory (DFT) calculations have predicted the presence of WPs which would be robust against lattice distortions in transition metal diphosphides (TMPs) WP$_2$ and MoP$_2$   \cite{Autes2016}. Similar to the TMDs, WP$_2$ and MoP$_2$ display both the XMR and the turn-on behaviour  \cite{Kumar2017}. The origin of these phenomena is not yet clear, with conflicting transport experiments supporting both the compensated \cite{Kumar2017, Schonemann2017} and uncompensated scenario \cite{Wang2017}.
To clarify this scenario, a momentum and energy-resolved technique like angle-resolved photoemission spectroscopy (ARPES) can be employed \cite{Damascelli2004}. However, the presence of both surface and bulk bands in the ARPES spectra of WP$_2$ complicate this approach.

Here, by combining ARPES experiments with DFT calculations, we reveal that this TMP is indeed a compensated semimetal in the bulk, albeit not at the surface. As ARPES is a highly surface-sensitive technique this result is illustrative of the need to disentangle surface and bulk states probed via ARPES in order to allow for a correct quantitative interpretation of the experimental data. 
In addition, we find that this semimetal hosts at least 4 WPs with coordinates in energy and momentum in qualitative agreement with the bulk DFT calculations. The bulk band structure is found to be unchanged over a wide range of temperature, demonstrative of the WPs insensitivity to temperature. 

The ARPES experiments were performed at the Merlin beamline at the Advanced Light Source (ALS). The samples were cleaved \textit{in situ} in the (010) plane and measured in a vacuum always better than $5\times 10^{-11}$ mbar. DFT calculations for the WP$_2$ Fermi surface in Fig. 1 (a, b) were performed with the WIEN2K software package  \cite{Wien2K}. Lattice constants and internal atomic positions were taken from the experimentally determined crystal structure \cite{Ruhl1983}. From these DFT calculations, we find that the bands crossing the Fermi level ($E_F)$ have nearly compensated hole ($n_h = 3.75 \times 10^{21}$ cm$^{-3}$) and electron ($n_e = 3.86 \times 10^{21}$ cm$^{-3}$) density, in agreement with previous results \cite{Kumar2017, Schonemann2017}. 

\begin{figure*}
\includegraphics[width=0.95\textwidth]{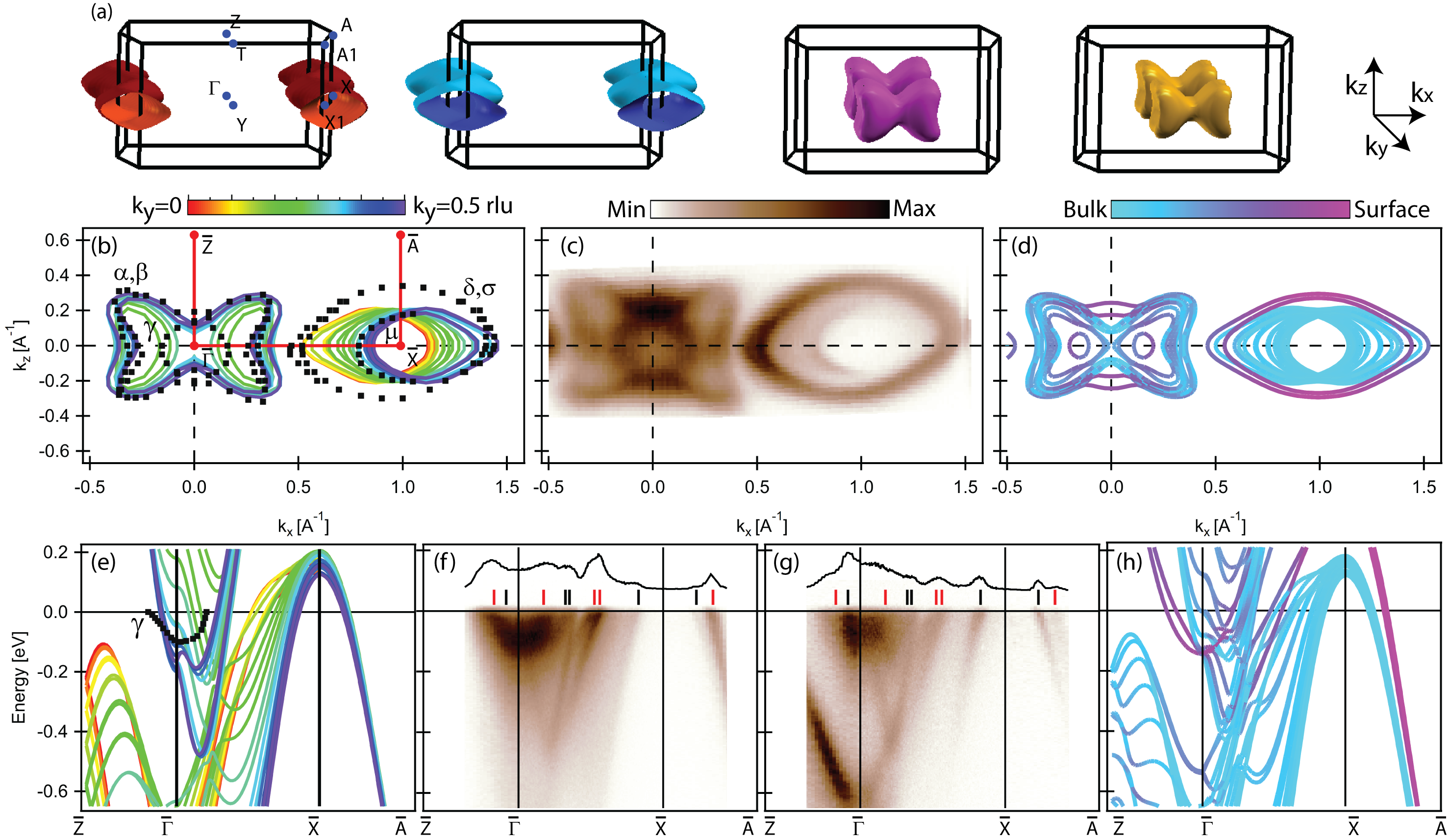}
\caption{DFT-calculations and ARPES intensity maps at $T=25$ K of WP$_2$. (a) Fermi surface sheets from DFT were plotted using XCRYSDEN visualization package \cite{xcrysden}.  (b) Bulk DFT calculations at  $E_F$ in the $k_x-k_z$ plane. Black squares indicate the $k_Fs$ extracted from the MDC peak position of the ARPES data in (c).(c) ARPES intensity maps at $E_F$ in the $k_x-k_z$ plane at $h \nu = 50$ eV. 
(d)  DFT band structure in the $k_x-k_z$ plane of a  WP$_2$-slab of 18 W-planes and 36 P-layers, for the relaxed structure. The colourscale represents the contribution from the top W layer (surface contribution) and from the layers below (bulk contribution).
(e) Bulk band structure along high symmetry lines [red path in panel (b)]. Black squares indicate the $k_F$ extracted from the MDC peak position of the data in (f).
(f), (g) ARPES intensity maps along high symmetry lines for $p-$ and $s-$polarized light, respectively. MDCs at the $E_F$ are shown in the inset. Red (Black) vertical lines identify surface (bulk) peaks in the MDCs. (h) Slab DFT band structure along high symmetry lines.}
\end{figure*}

\begin{figure}[b]
\includegraphics[width=0.5\textwidth]{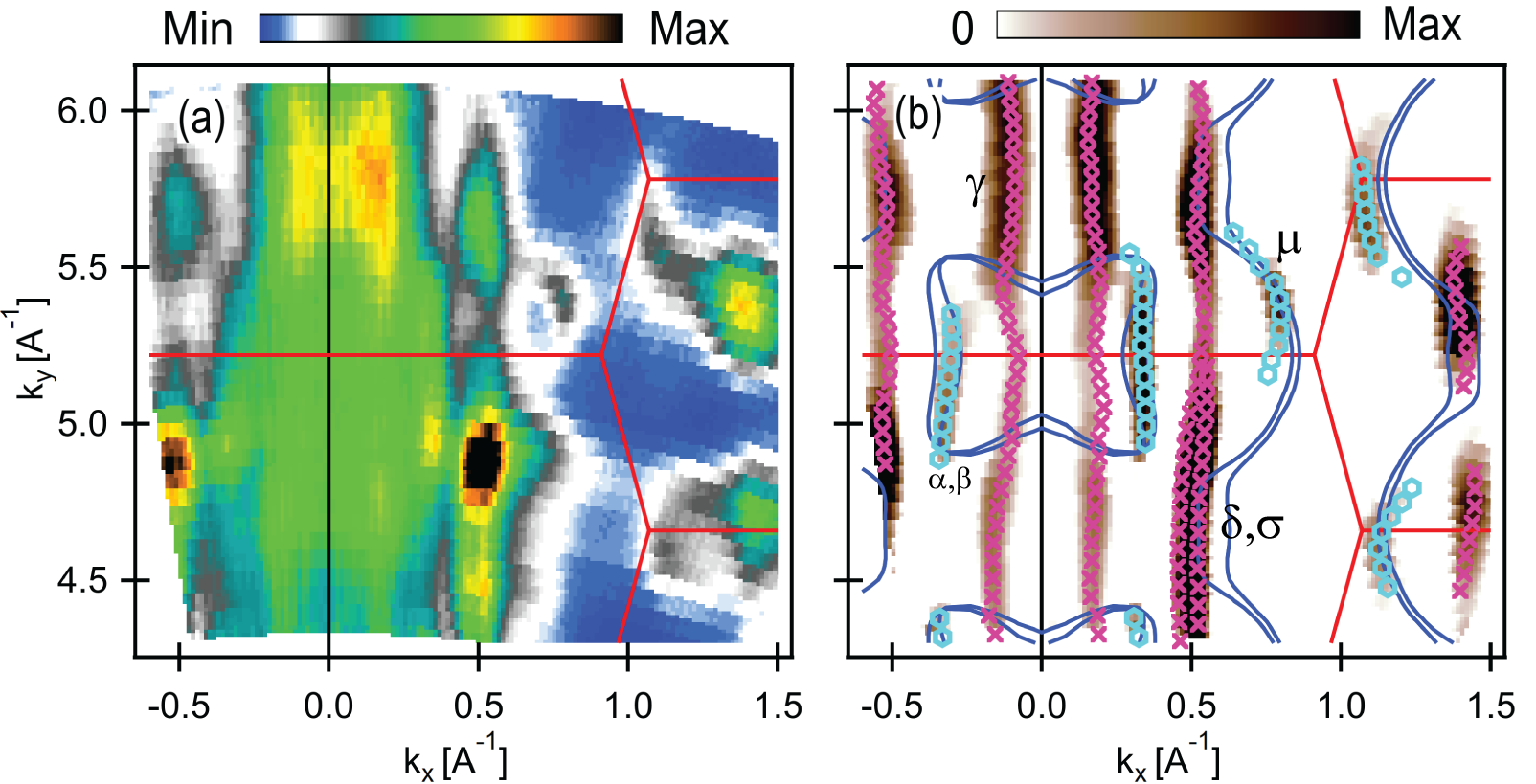}
\caption{ (a) ARPES intensity maps at $E_F$ in the $k_x-k_y$ plane measured by tuning the photon energy from 50 to 120 eV. (b) MDC second derivative intensity plot, MDC peak positions for bulk (azure hexagons) and surface bands (pink crosses)  of the ARPES data in (a). Bulk-DFT calculations (black lines) are superimposed with the data. Red lines denote the boundaries of the BZ. }\label{SS_DFT}
\end{figure}

In Fig. 1 (c) we plot the ARPES-derived Fermi surface in the $k_x-k_z$ plane, as measured with $h \nu = 50$ eV photon energy. Under the free-electron final-state approximation (FEFSA)  and an inner potential $E_0=26 \text{ eV,}$ this corresponds to $k_\perp=k_y= 0.32$ r.l.u. in the reduced zone scheme. Indeed, comparison between the ARPES-measured and DFT-calculated Fermi surface reveals that the largest electron pockets at $\overline{\Gamma}$ [$\alpha, \beta$ bands in Fig. 1(b)], and the smaller hole-pocket at $\overline{X}$ ($\mu$) are in agreement for $k_y = 0.32$ r.l.u. 
However, additional bands are observed ($\gamma$, $\delta$ and $\sigma$) with no clear counterpart in the bulk-DFT calculations.
In particular the hole $\delta$ and $\sigma$ pocket areas in the $k_x-k_z$ plane are $4.49 \times 10^{15}$ cm$^{-2}$ and $4.17 \times 10^{15}$ cm$^{-2}$, respectively.  These results are significantly larger than any cross sectional DFT Fermi surface centered around $\overline{X}$, e.g. $A^{Bulk}_{MAX}=1.83 \times 10^{15}$ cm$^{-2}$ at $k_y=\pi/c$. 

\begin{figure*}
\includegraphics[width=1\textwidth]{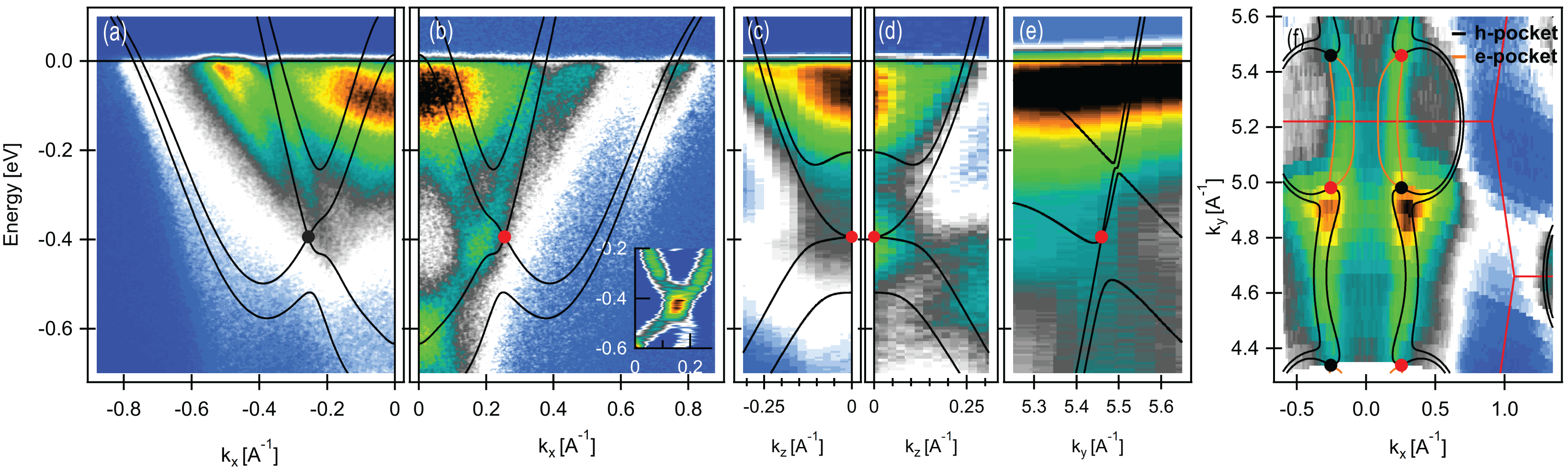}
\caption{ARPES intensity maps. (a) ARPES intensity map at one of the Weyl points.   (a),(b) ARPES intensity maps along $(k_x, 4.344, 0)$ for $p-$ and $s-$polarized incident light, respectively. Inset in (b) is the MDC second derivative intensity plot. (c), (d)  ARPES intensity map at the Weyl point along $(0.19, 4.344, k_z)$ for $p-$ and $s-$polarized incident light, respectively. (e) ARPES intensity map at the Weyl point along $(0.19, k_y, 0)$ for $p-$polarized incident light.  Bulk-DFT calculations at the Weyl point superimposed with the data. (f) ARPES intensity maps at $E_b=400$ meV in the $k_x-k_y$ plane obtained by tuning the photon energy from 50 to 120 eV. DFT calculations (black and orange lines) are superimposed with the data. Red lines denote the boundaries of the BZ.  }
\end{figure*}
%

Previous ARPES results \cite{Kumar2017} have suggested that this discrepancy may come as a result of substantial $k_\perp$-integration \cite{Strocov2003}: the combination of finite mean-free path for the photoemitted electron with the small size of the Brillouin zone (BZ) along $k_y$ could allow for the superposition of photoemission intensity from states projected along this axis of the Brillouin zone. 
The size of  the hole pockets along $k_x$ ($\overline{\Gamma}-\overline{X}$) are indeed consistent with what is expected from a projection of the DFT calculation along $k_\perp$. However, the $k_F$  along the $k_z$ direction ($\overline{X}-\overline{A}$) extracted from the momentum distribution curve (MDC) peaks is larger than that calculated for any $k_\perp$ and therefore a description based only on $k_\perp$ integration is insufficient to explain the ARPES spectra. Deviations between the calculation and the ARPES data are also present at the zone centre. A small circular FS-sheet, with an area of $A = 0.75 \times 10^{15}$ cm$^{-2}$, is observed around $\overline{\Gamma}$. This band, which we label as $\gamma$,  is electron-like [see Fig. 1 (f)] and is not present in the bulk DFT calculation for any $k_\perp$, as show in Fig. 1 (e). Its intensity is reduced when using $s-$polarized light [see Fig. 1 (g)].

With these inconsistencies between the photoemission and DFT-derived band structure established, we seek to determine their origin. Anticipating the possible role of surface states which may complicate the ARPES spectra, we performed DFT calculations for a P-terminated slab of 18 W-layers and 36 P-layers. The results are shown in Fig. 1 (d) and (h).  The colourscale represents the contribution of the top W layer (surface contribution) and the layers below (bulk contribution). To simulate the effect of cleaving a bulk crystal as done in the experiment, the slab crystal structure was relaxed to achieve force minimization \cite{Supplementary}. Beside the states corresponding to the projected bulk states [cyan bands in Fig.1 (d) and (h)], 4 additional bands of predominant top-most surface origin cross $E_F$ [pink bands in Fig. (d) and (h)]. These surface states (SS) are in qualitative agreement with the bands $\gamma$, $\delta$ and $\sigma$ described above.
We emphasize here that these SS are closed and therefore do not represent the Fermi arcs characteristic of Weyl semimetals. This result is consistent with the fact that the projected Weyl charge is 0 on the (010) surface and no Fermi arcs are expected to form \cite{Wan2011}.

To further confirm the surface origin of these states, we performed ARPES measurements at variable photon energy to explore their variation along the $\overline{\Gamma} - \overline{X}$ direction. As a surface state is by definition restricted to the two-dimensional surface Brillouin zone, it should have no dispersion along this direction of momentum space.
Figure 2 (a) and (b) show the ARPES intensity and its MDC second derivative at $E_F$ in the $k_z=0$ plane. Conversion from $h\nu$ to $k_\perp(=k_y)$ values was performed by using the FEFSA with an inner potential $E_0=26$ eV. 
The $h\nu$-dependent maps confirm that the bands identified as originating from the bulk are three dimensional [cyan hexagons in Fig. 2 (b)] and in quantitative agreement with our bulk-DFT calculations. In contrast to this, the additional bands observed in ARPES are highly two-dimensional (2D), and exhibit negligible dispersion in the direction perpendicular to the cleavage plane [pink crosses in Fig. 2 (b)] in confirmation of their designation as surface states.
%

\begin{figure*}
\includegraphics[width=1\textwidth]{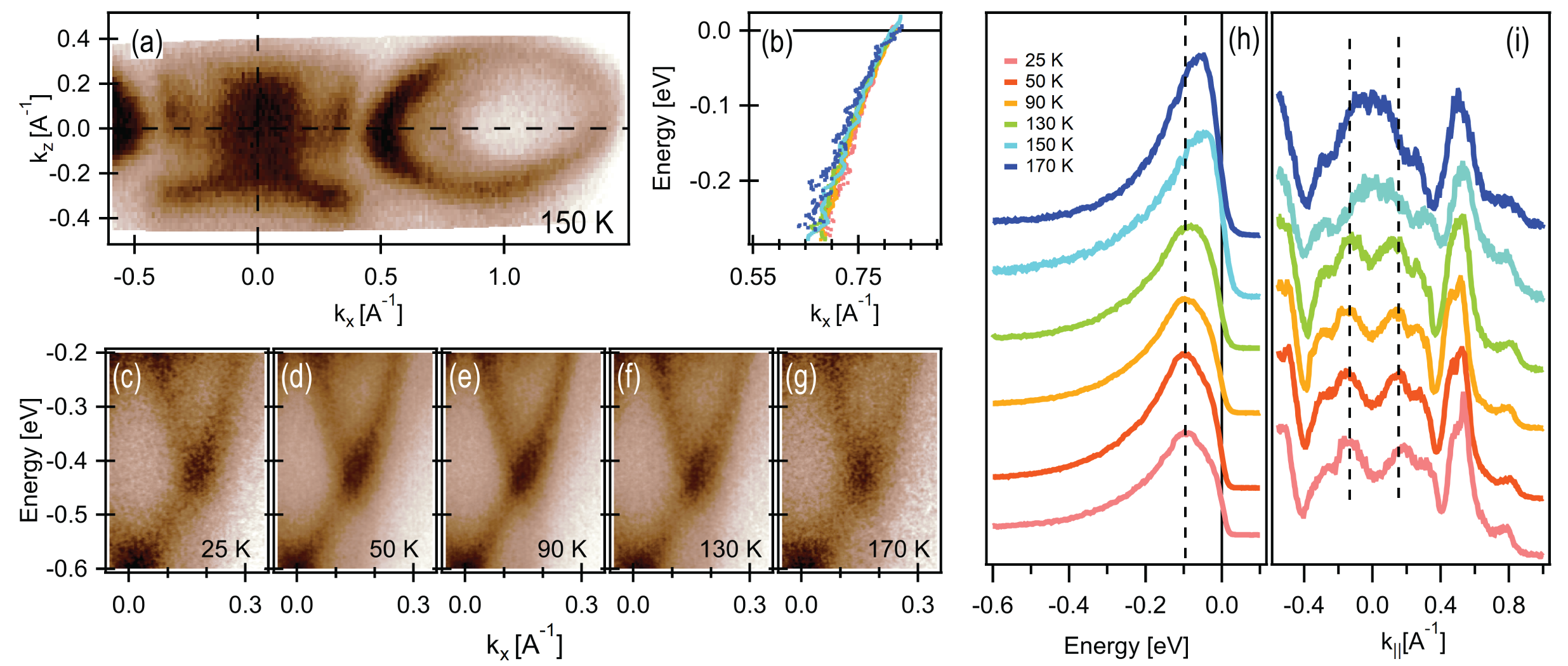}
\caption{Temperature dependence of ARPES band structure. (a) ARPES intensity maps at $E_F$ in the $k_x-k_z$ plane at $h \nu = 50$ eV collected at $T=150$ K. (b) $T$-dependent dispersion of bulk-band $\mu$ [defined in Fig. 1 (b)] close to $\overline{X}$ extracted from MDC peaks. (c)-(g) ARPES intensity maps along $\overline{\Gamma}-\overline{X}$ for $s-$polarized light at various temperatures. (h) EDCs at  $\overline{\Gamma}$ as a function of temperature. (i) MDCs along $\overline{\Gamma}-\overline{X}$ at $E_F$ as a function of temperature.}
\end{figure*}

Having disentangled bulk and surface states in WP$_2$,  we now examine how the bulk bands compare against the positions of the WPs predicted from DFT calculations. In our DFT calculation the 2 quartets of WPs are located at
\{$\boldsymbol{k}^{\pm W(1)} =( 0.2552, \pm 0.3226, 0) \text{\normalfont\AA}^{-1}$, $E_b= 395$ meV \}  and \{$\boldsymbol{k}^{\pm W(2)} =(0.2471,\pm 0.2882, 0 ) \text{\normalfont\AA}^{-1}, E_b=255$ meV \}, respectively.
Figure 3 demonstrates that, even in the presence of a non-negligible $k_\perp$-broadening, our ARPES measurements resolve 4 Weyl points located about 400 meV below $E_F$.  A cone-like dispersion in qualitative agreement with our calculations can be observed along the $k_x$ and $k_y$ directions centred at $\boldsymbol{k}^{WP}$= $(0.19, 0.32, 0)$ \AA$^{-1}$, in the reduced zone scheme (in the extended zone scheme used in Fig. 3 (e)-(f) this corresponds to  $k_y^{WP}= 4.344, 4.981$ and $5.4591 $ \AA$^{-1}$).
Given the finite $k_\perp$ resolution and presence of multiple bands in a narrow $k$ range along this direction, the cone is not easily distinguished along $k_y$. Ultimately, the ARPES intensity map shown in Fig. 3 (e) is in qualitative agreement with the DFT bulk calculation for energies below the SS, i.e., for $E_b<-250$ meV. As one would anticipate for a type-II semimetal, the Weyl points coincide with the intersection of hole-like and electron-like constant energy contours in the $k_x$-$k_y$ plane [see Fig. 3 (f)].
It is important to note that a further 4 WPs are not observed in this experiment. However, their anticipated proximity in energy and momentum to the SS, which themselves produce exceptionally high photoemission intensity, may obfuscate these additional WPs. While we can not easily distinguish these features in the bulk spectra, these results do not themselves rule out the presence of the 4 additional WPs predicted by DFT.

The stability of the band structure against small variations of the crystal lattice is an important issue in the TMDs. In WTe$_2$ for example, strain can shift neighboring WPs towards each other until they touch and annihilate due to their opposite chiral charge \cite{Bruno2016}. Dramatic changes in the band structure of WTe$_2$ below 50 K, observed in ARPES \cite{Pletikosi2014}, time resolved reflectivity \cite{Dai2015} and transport \cite{Thoutam2015} have been put forth to explain the turn-on behaviour and  disappearance of XMR at high $T$. 
In WP$_2$ however, the neighbouring WPs have the same chirality and the DFT electronic structure is insensitive to small variations in the crystal structure \cite{Autes2016}. Consequently, the WPs here are predicted to be inert to similar variations of the crystal structure which would coincide with the application of strain or change of temperature. 

This prediction is confirmed by our measurements. While in WTe$_2$ the hole pocket shrinks by about 50$\%$ when the temperature is raised from 25 to 100 K \cite{Pletikosi2014} and completely disappears around 160 K \cite{Wu2015}, in  WP$_2$ the hole pocket  area at the zone edge is unchanged between 25 and 150 K [see Fig. 4 (a)]. In particular, the band dispersion of the bulk band $\mu$ extracted from the MDCs peaks [panel (b) in Fig. 4] does not show any substantive change. This holds for both $k_F$ and Fermi velocity, as measured up to 170 K. Similarly, the dispersion at the Weyl points is unchanged over this temperature range [panel (c)-(g) in Fig. 4].  
At the zone center, only the SS $\gamma$ shows a non-negligible modification moving towards $E_F$ above 90 K [see Fig (h) and (i)]. This observation is consistent with our slab-DFT calculation, where the SS is indeed expected to be more sensitive to small changes in the crystal structure, even moving above $E_F$ for small values of strain \cite{Supplementary}. 
The remaining bulk-bands at the zone center are also robust up to 170 K, as demonstrated by the $T$-dependence of the MDCs [in Fig. 4 (i)].

Overall our results indicate that WP$_2$ differs substantially from WTe$_2$, where the Fermi surface areas, and thus the carrier densities $n_e$ and $n_h$, 
are 
highly renormalized at high temperatures. 
As demonstrated by our combined ARPES-DFT analysis (Fig.1), and the ARPES study vs temperature (Fig. 4), in WP$_2$ the bulk carrier density for electrons and holes are nearly perfectly compensated and $T$-independent. These results are then consistent with the prediction that the XMR in WP$_2$ stems from electron-hole compensation. However, the turn-on behaviour cannot be explained as a result of changes to the band structure induced by temperature, as it has been proposed for WTe$_2$ \cite{Pletikosi2014, Dai2015, Thoutam2015}.  In contrast to this, these results are more suggestive of theories based on strong $T$-dependence of the carrier mobility \cite{Wang2015, Pei2017} and/or band structure modifications induced by the magnetic field itself \cite{Trescher2017}. Finally, we note that the surface-to-bulk progression of the electronic structure observed here, and its impact in assessing the presence or lack thereof of carrier compensation in WP$_2$, might be of general relevance to other TMPs or TMDs, and should therefore be taken into account in a quantitative interpretation of ARPES data.

To bring clarity to the origins of XMR in TMPs, we have carried out a combined DFT and ARPES study of WP$_2$.  Based on bulk- and slab-DFT calculations, we have identified surface and bulk contributions to the ARPES signal. ARPES intensity originating from bulk states is in agreement with DFT calculations, displaying the anticipated electron-hole compensation on the Fermi surface as well as at least 4 Weyl points. Furthermore, despite the opposite behaviour found in the related TMDs, the bulk electronic structure is found to be stable over a large range of temperature, placing strict constraints on the possible origin of the turn-on behaviour for XMR in this material.

\begin{acknowledgments}
This research was undertaken thanks in part to funding from the Max Planck-UBC-UTokyo Centre for Quantum Materials and the Canada First Research Excellence Fund, Quantum Materials and Future Technologies Program. The work at UBC was supported by the Killam, Alfred P. Sloan, and Natural Sciences and Engineering Research Council of Canada (NSERC) Steacie Memorial Fellowships (A.D.), the Alexander von Humboldt
Fellowship (A.D.), the Canada Research Chairs Program (A.D.), NSERC, Canada Foundation for Innovation (CFI), and CIFAR Quantum Materials Program. E.R. acknowledges support from the Swiss National Science Foundation (SNSF) grant no. P300P2$\_$164649. 
\end{acknowledgments}


\end{document}